\documentclass[preprint,prd,nofootinbib]{revtex4}
\pdfoutput=1
\def\gsim{ \lower .75ex \hbox{$\sim$} \llap{\raise .27ex
\hbox{$>$}} }
\def\lsim{ \lower .75ex \hbox{$\sim$} \llap{\raise .27ex
\hbox{$<$}} }

\usepackage{graphicx}
\usepackage{dcolumn}
\usepackage{bm}
\usepackage{graphicx}

\usepackage{amssymb}
\usepackage{bbm}

\begin{document}

\title{Boom and Bust Inflation: \\ a Graceful Exit via Compact Extra Dimensions}

\author{Adam R. Brown}

\affiliation{Physics Department, Columbia University, New York, NY 10027,
USA}

\begin{abstract}

A model of inflation is proposed in which compact extra dimensions allow a graceful exit without recourse to flat potentials or super-Planckian field values. Though bubbles of true vacuum are too sparse to uniformly reheat the Universe by colliding with each other, a compact dimension enables a single bubble  to uniformly reheat by colliding with itself. This mechanism, which generates an approximately scale invariant perturbation spectrum, requires that inflation be driven by a bulk field, that vacuum decay be slow, and that the extra dimension be at least a hundred times larger than the false vacuum Hubble length.

\end{abstract}

\maketitle

\section{Introduction} 
Many theories posit compact extra dimensions for reasons unrelated to cosmology \cite{ued,add,rs} --- this paper will show that in doing so you may get a free bonus: a graceful exit from inflation without recourse to flat potentials or super-Planckian field values, and with a mechanism for generating approximately scale invariant perturbations. 

The Universe is very flat. The spatial curvature radius, $l_{curv}  (\textrm{today})$, is observed to be no less than ten times the current Hubble length, $H_0^{-1}$ \cite{wmap}. Using the entropy density, $S$, we can construct the dimensionless (and, for adiabatic evolution, conserved) quantity
\begin{equation}
l_{curv} \,  S^{1/3} > e^{70}. \label{entropy}
\end{equation}
This number is huge. Inflation explains why.

\begin{figure}[h] 
\centering
   \includegraphics[width=.4\textwidth]{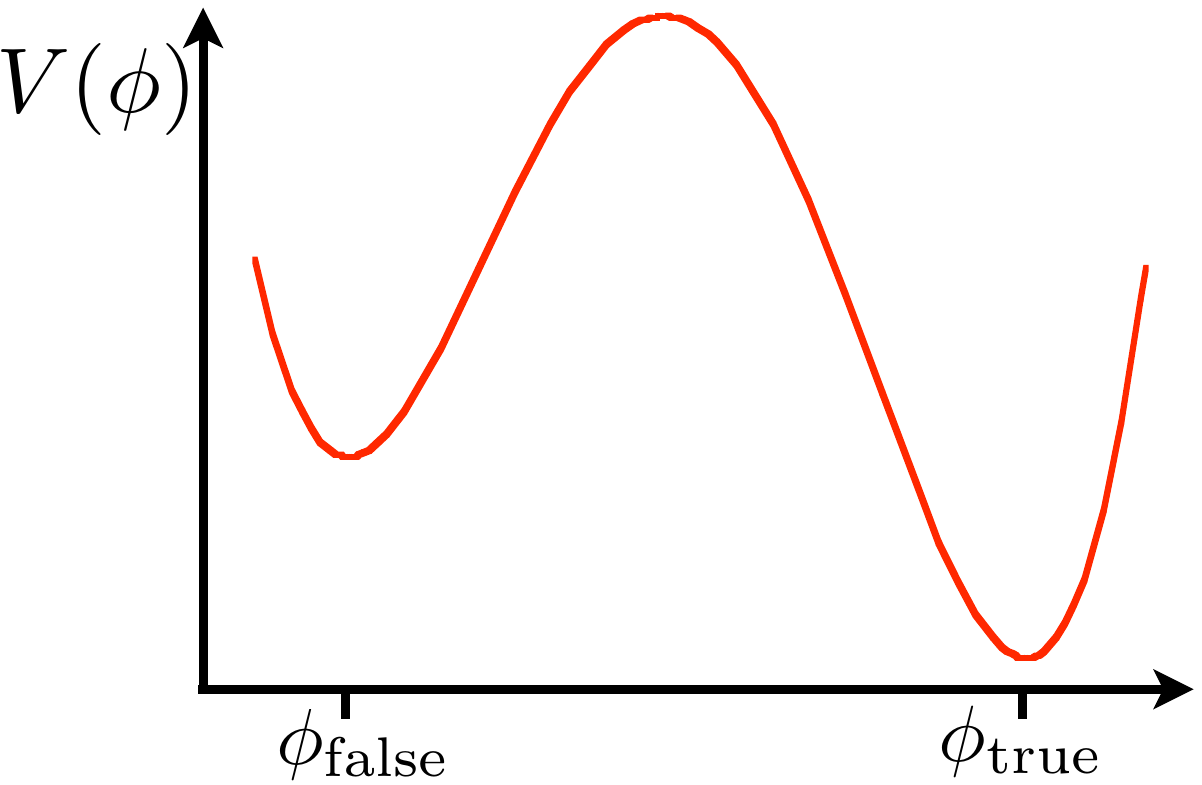} 
\centering
   \caption{The inflaton potential both for old inflation and for boom and bust inflation.}
\label{fig1}
\end{figure}

In old inflation \cite{guth}, the Universe gets temporally stuck in a false vacuum during an early first order phase transition. While in this false vacuum the Universe (adiabatically) inflates, safely diluting primordial entropy and monopole density, and solving the horizon problem. Inflation ends, locally, when the false vacuum decays by nucleating bubbles of true vacuum \cite{coleman} and the Universe (nonadiabatically) reheats, locally, when many such bubbles collide with one another. But old inflation does not work: if inflation is to last sufficiently long to do its job then decay must be slow, so bubbles are sparse, collisions are rare, and percolation fails \cite{guth,gw}.

New inflation \cite{slowroll} does work. It instead recovers the observable Universe from a single bubble --- the interior of a bubble has the geometry of an open Friedmann-Robertson-Walker (FRW) Universe. New inflation adds an exceptionally flat segment on the true vacuum side of the potential barrier so that inflation continues after nucleation while the field rolls slowly down. This post-nucleation period of inflation continues until the field reheats at the bottom of the potential, by which time the curvature radius has grown sufficiently to satisfy Eq.~(\ref{entropy}).

In this paper we introduce a new model of inflation, which we call boom and bust inflation. Like old inflation it makes do with an unexceptional inflaton potential. Like old inflation it reheats via bubble collision. But like new inflation it recovers the visible Universe from a single bubble. To do so it borrows structure that has been independently invoked: compact extra dimensions\footnote{For other inflationary scenarios involving bubbles in extra dimensions, see \cite{TandB}.}.

\section{Self-intersecting bubbles} \label{self-intersecting}
In this section, as a proof of principle, we study the growth of a bubble in a particularly simple inflating space --- four-dimensional de Sitter spacetime plus a single flat non-inflating compact extra dimension, with the inflation driven by the false vacuum energy of a five-dimensional scalar field. Before the bubble nucleates, and outside of the bubble after it nucleates, the metric is
\begin{equation}
ds^2 = -d\tilde{t}^2 + e^{2H \tilde{t}} d\vec{x}^{\, 2}  + dy^2  \, \, , \label{metric}
\end{equation} 
where $y$ is periodically identified with period $d$, $H^{-1}$ is the false vacuum Hubble length, and $d \vec{x}^2 = d\tilde{r}^2 + \tilde{r}^2 d \theta^2 + \tilde{r}^2 \sin^2 \theta d \phi^2$. This space has the ten SO(4,1) de Sitter symmetries, as well as a single $y$-translation symmetry. The bubble breaks all five ``translation'' symmetries, leaving the six SO(3,1) rotation and ``boost'' symmetries that keep the nucleation point invariant. 
\begin{figure}[h!]
   \centering
   \includegraphics[width=.5\textwidth]{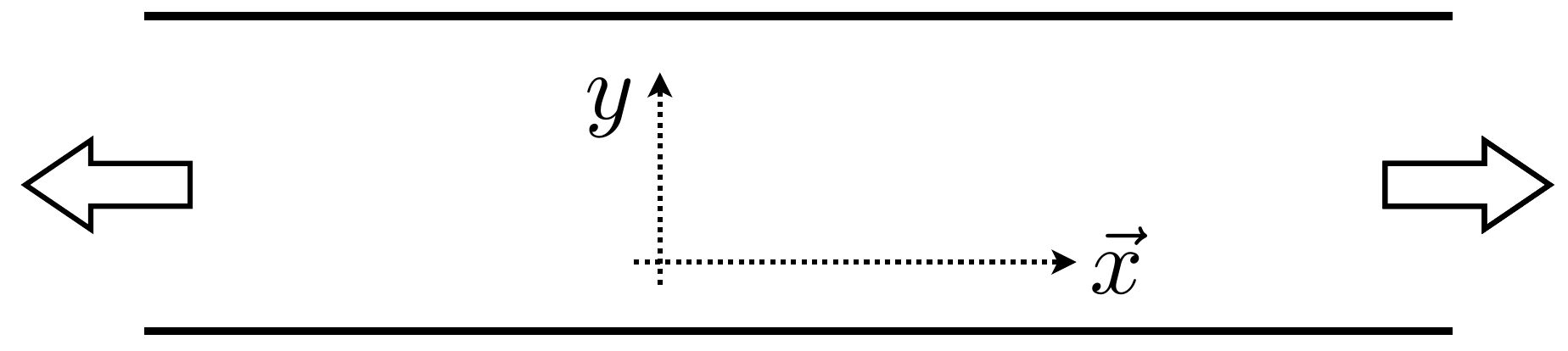} 
   \caption{A snapshot of the spacetime of Eq.~(\ref{metric}). Three of the spatial dimensions, $\vec{x}$, are inflating. The fourth spatial dimension, $y$, is fixed and has period $d$ --- the top and bottom lines (representing 3-surfaces) are identified.}
   \label{di2}
\end{figure}

We will, in this section, make three further assumptions: that decay is very rare, so that collisions with other bubbles can be safely neglected; that $\bar{R}$, the size of the bubble in the compact direction at nucleation, is much less than $d$ and $H^{-1}$, so that the bubble wall traces the light cone of its nucleation point; and that the radion mass is very large, so that $d$ is essentially fixed. With these assumptions we will study the growth of a bubble that nucleates at $\tilde{r} = \tilde{t} = y =0$, first for the case $d \ll H^{-1}$, which does not give a graceful exit, and then for the case $d \gg H^{-1}$, which does.

\subsection{Self-intersecting bubbles with $d \ll H^{-1}$}

\begin{figure}[h!] 
   \centering
   \includegraphics[width=.5\textwidth]{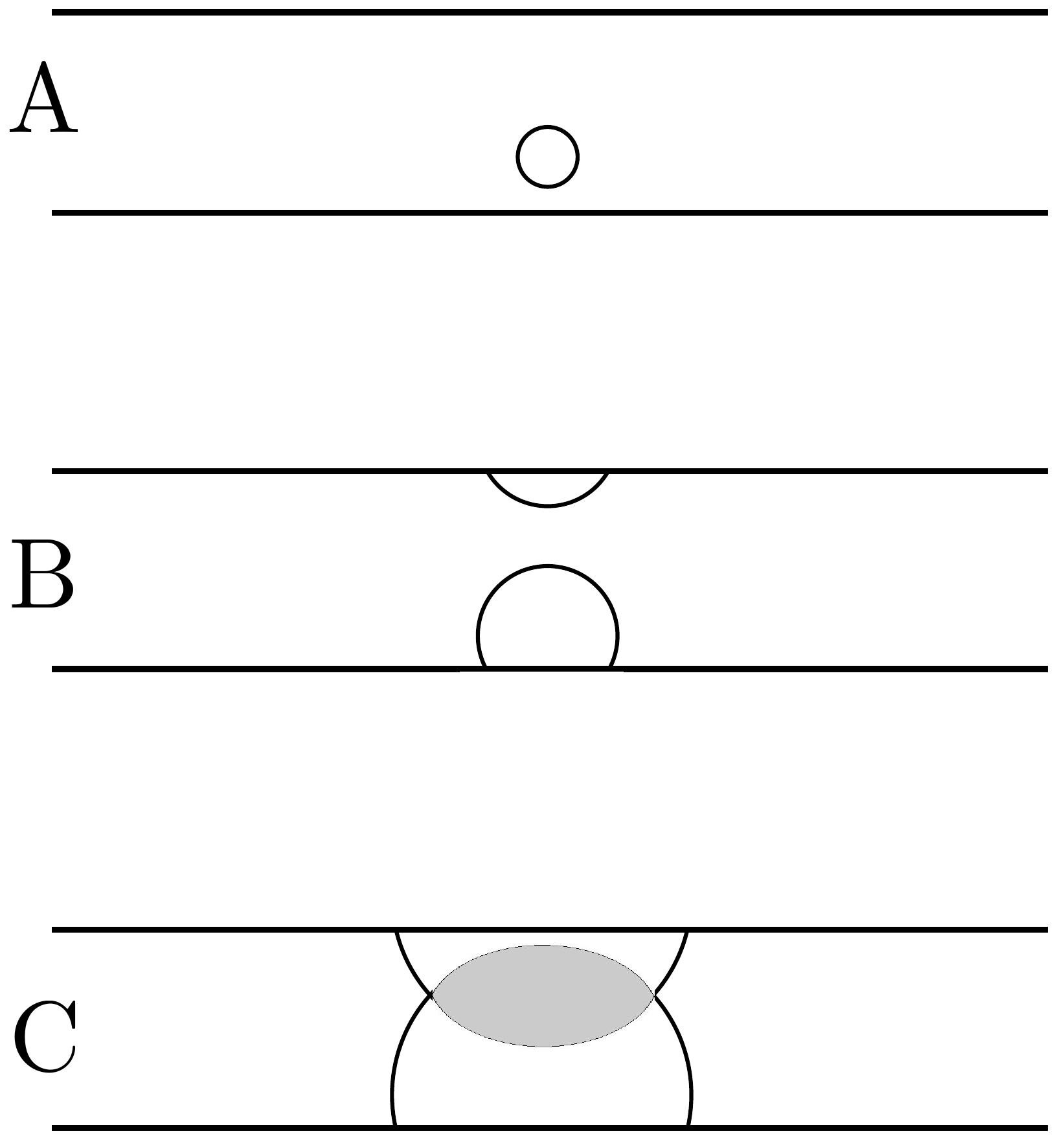} 
   \caption{Three snapshots at increasing values of $\tilde{t}$, for $d \ll H^{-1}$. (A) The bubble of true vacuum forms within the false vacuum. (B) The bubble grows; all the latent energy is in the bubble wall. (C) The bubble collides with itself, locally reheating the Universe.}
   \label{di3}
\end{figure}

For $d \ll H^{-1} $ we may ignore the inflation of the external space until the bubble is much larger than $d$. Just as in five-dimensional Minkowski space, the bubble grows spherically as a four-ball of true vacuum [its surface given by $-\tilde{t}^2 + \tilde{r}^2 + y^2 = 0$, up to corrections of order $H \tilde{t}$ (and $\bar{R}$)]. When the diameter of this four-ball reaches $d$, one part of the bubble wall ($\tilde{r}=0, \, y=\frac{d}{2}$) collides with another ($\tilde{r}=0, \, y=-\frac{d}{2}$) after winding around the compact dimension (see Fig.~\ref{di3}). This collision, and recollisions of the emitted radiation on subsequent laps, releases the bubble wall energy, reheating first the inflaton and then the Standard Model fields; very soon homogeneity in $y$ is recovered.

The constant-$\tilde{t}$ snapshots of Fig. 3 pick out $\tilde{r}=0$ as privileged --- the collision happens there first. But a boost along the $\vec{x}$ directions will result in a different time-slicing and a different point of ``first'' contact. Indeed, the SO(3,1) symmetry ensures that all points on the three-dimensional surface of bubble self-intersection are physically equivalent, and that this surface is a hyperboloid of constant (negative) curvature. Up to corrections of order $H \tilde{t}$, the surface is given by
\begin{equation}
y = \frac{d}{2}, \ \ \ \tilde{r}^2 - \tilde{t}^2 = -(\frac{d}{2})^2, \ \ \ \tilde{t}>0 \, .
\end{equation}
\begin{figure}[h!]
   \centering
   \includegraphics[width=.5\textwidth]{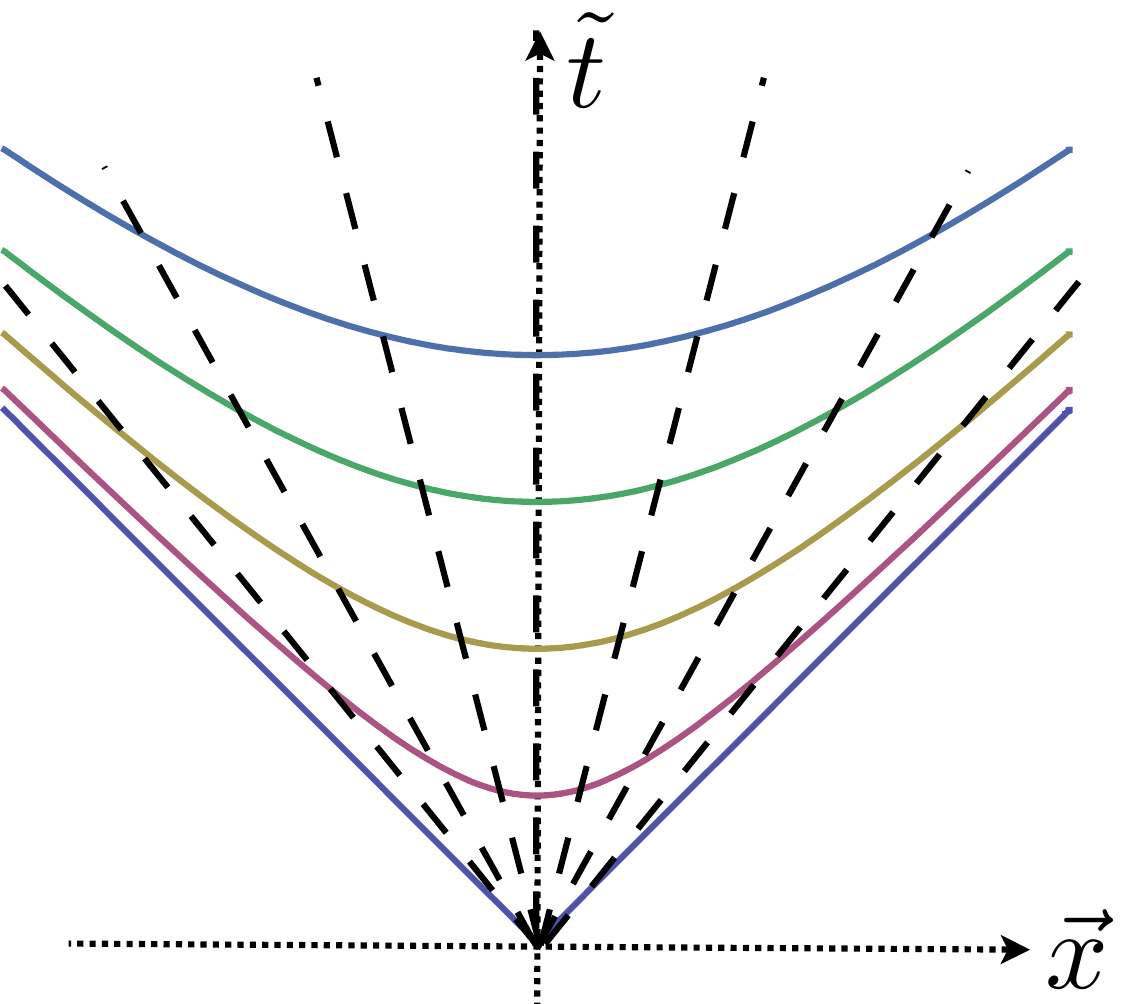} 
   \caption{Milne coordinates foliate four-dimensional Minkowski space with hyperboloids (solid lines of constant $r$). Geodesics through the origin follow dashed lines of constant $t$.}
   \label{Milne}
\end{figure}

We can choose coordinates that respect the SO(3,1) symmetries of the nucleated bubble, so that the Universe is spatially homogeneous and the collision happens everywhere simultaneously. These are Milne coordinates, defined by $t = \sqrt{\tilde{t}^2 - \tilde{r}^2} $, $r = {\tilde{r}}/{\sqrt{\tilde{t}^2 - \tilde{r}^2}}$, which foliate four-dimensional Minkowski spacetime with hyperboloids to give a metric
\begin{equation} \label{eq:milne}
ds^2 = - d t^2  + t^2 \Bigl[ \frac{d r^2 }{1 + r^2} + r^2  (d \theta^2 + \sin^2 \theta d \phi^2 ) \Bigl] + dy^2 \ . 
\end{equation} 
These coordinates cover the causal future of $t = 0$. Light rays emanating from the nucleation point at $y = t  = 0$ follow
\begin{equation}
y = t,  \ \ \ \ \ \textrm{with constant } r, \ \theta, \ \phi,
\end{equation}
so the three-hyperboloid of bubble self-intersection is
\begin{equation}
y = t =  \frac{ d}{2}  \ ,
\end{equation}
which implies 
\begin{equation}
l_{curv}(\textrm{collision}) \sim \frac{ d}{2}. \label{curv}
\end{equation}
The collision surface becomes the reheating surface (approximately --- some of the energy makes multiple laps before reheating), so this curvature radius is (approximately) inherited by the resultant open FRW Universe. The subsequent expansion stretches this to
\begin{equation}
l_{curv}(\textrm{today}) \sim \frac{S(\textrm{reheating})^{\frac{1}{3}}}{S(\textrm{today})^{\frac{1}{3}}} \frac{d}{2} \, .
\end{equation}
Even if $S(\textrm{reheating})^{{1}/{3}}$ is as great as the five-dimensional Planck mass $M_{5\textrm{\scriptsize Pl}} = M_{4\textrm{\scriptsize Pl}}^{{2}/{3}} \, d^{- {1}/{3}}$, and even if the extra dimension is so large as to saturate the E\"{o}tv\"{o}s upper bound  of $d=50\mu m$ \cite{eotvos}, then today's spatial curvature radius would still be only ten light years: far too small.

The nucleation and self-destruction of the bubble reintroduces a flatness problem, one of the very problems inflation was invented to resolve. What we need is a new period of inflation to solve this problem. This is provided if $d  \gg H^{-1}$.

\subsection{Self-intersecting bubbles with $d \gg H^{-1}$}
In this subsection we will exclusively discuss the bubble wall as viewed from the region outside the bubble, which inflates per Eq.~(\ref{metric}). We will discuss the region inside the bubble, which does not inflate, in section~\ref{sec:Outlook}. For $d \gg H^{-1}$ the bubble wall's growth significantly deviates from spherical: the inflation redshifts the wall's angle of incidence to the $y$ axis, so that $\frac{d \tilde{r}}{d y} (\tilde{t}) = e^{- 2 H \tilde{t}} \frac{d \tilde{r}}{d y} (0)$, and stretches the bubble in the $\vec{x}$ directions, making the bubble exponentially oblate.

\begin{figure}[h!]
   \centering
   \includegraphics[width=.5\textwidth]{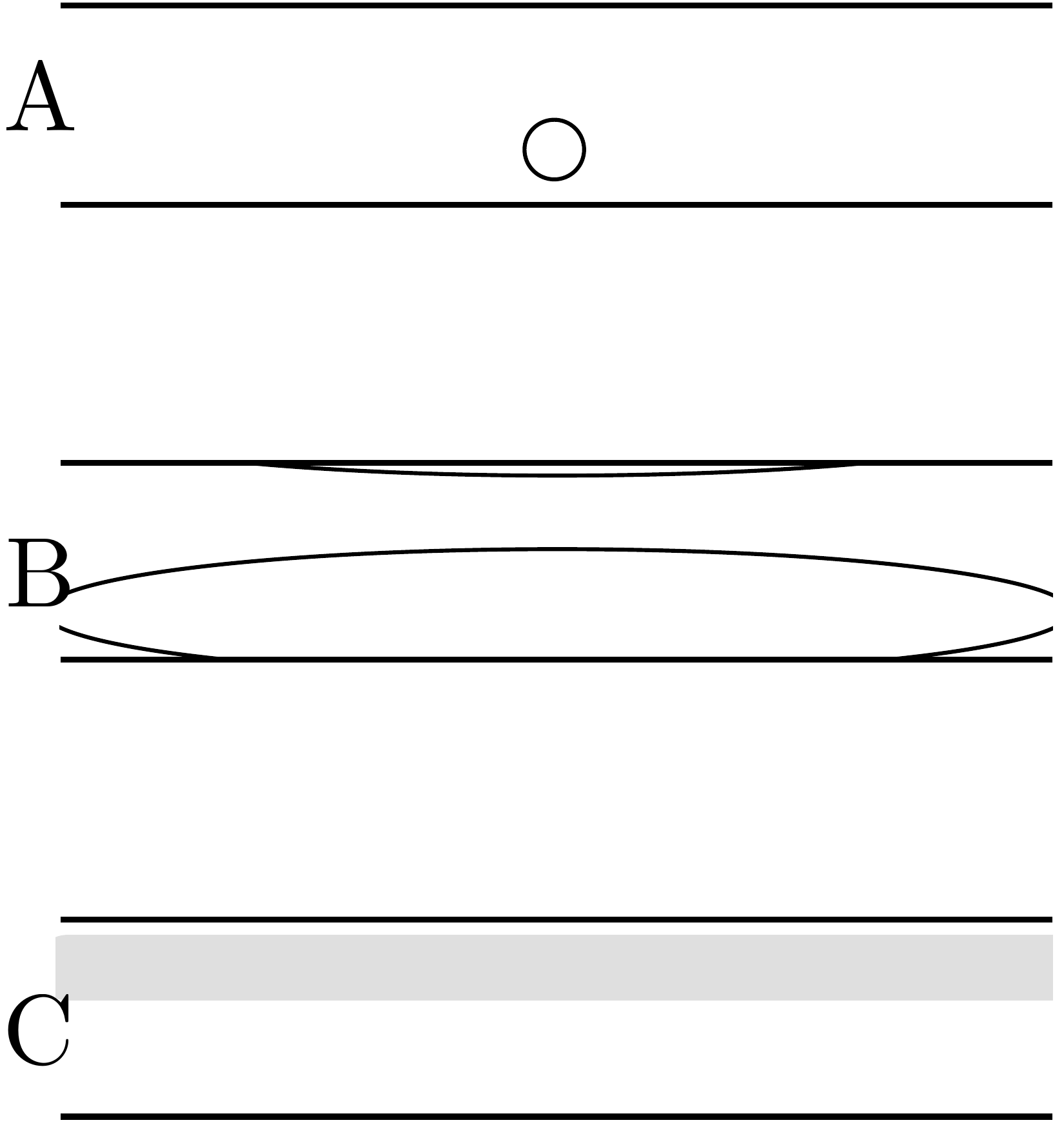} 
   \caption{Three snapshots at increasing values of $\tilde{t}$ for $d \gg H^{-1}$. (A) The bubble of true vacuum forms within the false vacuum. (B) As the bubble grows, the inflation in the $\vec{x}$ directions makes the bubble exponentially oblate (it ``booms''). (C) The bubble collides with itself (it ``busts'').}
   \label{di4}
\end{figure}

Just as before, the SO(3,1) symmetry ensures that the three-dimensional surface of self-intersection is a spatial hyperboloid, only now the curvature radius is exponentially large. To show this, we again change coordinates from the planar foliation of Eq.~(\ref{metric}), for which the collision happens at different times in different places, to a hyperbolic foliation, for which the collision is everywhere simultaneous. If $\cosh H t = \cosh H \tilde{t} - \frac{1}{2} H^2 \tilde{r}^2 e^{H \tilde{t}}$ and $r = H \tilde{r} \left[ \frac{1}{4} (1 + H^2 \tilde{r}^2 - e^{-2 H \tilde{t}} )^2 - (H \tilde{r})^2 \right] ^{-1/2} $, then
\begin{equation} \label{eq:milnedesitter}
ds^2 = - d t^2  + H^{-2} \sinh ^2 H t \left[ \frac{d r^2}{1 +r^2} + r^2 (d \theta^2 + \sin^2 \theta d \phi^2 ) \right] + dy^2 \ . 
\end{equation}
These coordinates cover the causal future of $t = 0$, but are not, of course, to be trusted inside the bubble, which is the causal future of the nucleation point at $y = t = 0$. Light rays emanating from the nucleation point again follow
\begin{equation}
y = t,  \ \ \ \ \ \textrm{with constant } r, \ \theta, \ \phi,
\end{equation}
so the three-hyperboloid of self-intersection is again
\begin{equation}
y = t = \frac{d}{2}, \label{bighyperboloid}
\end{equation}
which now implies
\begin{equation}
l_{curv}(\textrm{collision}) \sim H^{-1} \sinh \frac{ H d}{2}. \label{bigcurv}
\end{equation}
By considering the region outside the bubble we have shown that while the wall slowly advances in the $y$ direction, the inflation in the large directions so stretches the bubble as to make the collision surface exponentially flat. But since this flatness is an  intrinsic property of the collision hyperboloid, observers on the bubble interior must agree.

For $d \gg H^{-1}$ the period of inflation between bubble nucleation and self-annihilation solves the flatness problem and ensures a graceful exit from inflation.

\section{Conditions} \label{conditions}

Hypothesised extra dimensions come in many shapes and sizes. To realise the graceful exit of the last section, a compact-extra-dimensional inflationary theory must satisfy three conditions, over and above those conditions (such as radion stabilisation) that any viable theory must already satisfy. 

First, inflation must be driven by a bulk field. If the inflaton is inherently four-dimensional then the bubble cannot self-annihilate and this mechanism cannot work. 

Second, the five-dimensional decay rate, $\Gamma_5$, must be small enough that the sections of the bubble wall that formed the visible Universe likely encountered no other bubble: 
\begin{equation} \label{decay}
\Gamma _5 < H^2 H_0^3 \frac{S(\textrm{reheating})}{S(\textrm{today})} \ .
\end{equation}
Fortunately vacuum decay, as a non-perturbative effect, is generically exponentially suppressed. If this condition is satisfied, then so too will be the condition for eternal inflation, $\Gamma_5 < d^{-1} H^4$: beyond the visible Universe there will forever be regions of true vacuum that have not been consumed by bubbles of false vacuum and continue to inflate.

Third, the extra dimension must be large enough to reproduce the observed flatness. A finite $\bar{R}$ shortens the delay between nucleation and collision by reducing the $y$-distance the walls must travel; it lengthens the delay by slowing the walls [see Eq.~(\ref{speed})]. Conservatively including just the first effect, and combining Eqs.~(\ref{entropy}) and (\ref{bigcurv}), 
yields
\begin{equation}
d  >  \bar{R} + 2 H^{-1} \log \frac{2 \, e^{70}  \, H }{S(\textrm{reheating})^{\frac{1}{3}}} \, \, .  \label{radionconstraint}
\end{equation}
If $\bar{R}$ does not significantly exceed $H^{-1}$, this amounts to a single condition, approximately $d \, \gsim \, 100H^{-1}$. 
Of our three conditions, this one is the most restrictive. There is no a priori reason to expect $d$ to be greater than $H^{-1}$, and yet we require it to be over a hundred times greater; worse, for many radion stabilisation mechanisms this condition will be technically unnatural. Just as new inflation requires an atypically flat potential, so boom and bust inflation requires an atypically large extra dimension.
It is notable that many popular theories of extra dimensions already require atypically large extra dimensions: ADD with two extra dimensions requires $d \sim 10^{14}M_{\star}^{-1} $, where $M_{\star}$ is the electroweak/Planck scale \cite{add}; RS1 requires $d \sim 30 k^{-1}$ where $k^{-1}$ is the five-dimensional warping length \cite{rs}.  Because of this condition we will need to take extra care that any Kaluza-Klein gravitons produced during reheating do not disrupt standard cosmology \cite{KKgraviton}.

\section{Outlook} \label{sec:Outlook}

The next step should be to realise boom and bust inflation within an explicit ultra violet model. This will allow us to calculate two things. 

First, the inside story. The interior of the bubble cannot be five-dimensional Minkowski spacetime. If it were, then the interior metric would be $ds^2 = d\hat{t}^2 + d \hat{r}^2 + \hat{r}^2 d \Omega_2^2 + d \hat{y}^2$ (where constant $\hat{y}$ slices align with constant $y$ slices); the bubble wall, viewed from the inside, would grow spherically as $-\hat{t}^2 + \hat{r}^2 + \hat{y}^2 = 0$ (up to some small finite-nucleation-size corrections); since the curvature radius of the bubble collision surface is the same whether measured from within the bubble [Eq.~(\ref{curv})] or without [Eq.~(\ref{bigcurv})], we would have
\begin{equation}
\frac{\hat{y}_{\textrm{coll}}}{2} = H^{-1} \sinh \frac{H d}{2};
\end{equation}
and so the extra dimension would grow exponentially large. Therefore whatever stabilises the radion must also deform the interior away from five-dimensional Minkowski. An explicit UV model will allow us to calculate the interior metric as well as the effect of extra-dimensional warping.

Second, perturbations. A modern inflationary theory is not expected merely to flatten, gracefully exit and be forgotten; it is also expected to generate primordial perturbations. We can illustrate one promising source by highlighting the similarities between new inflation and boom and bust inflation, and the analogous roles played by $\phi(x)$ in the former and by $y(x)$, the location of the bubble wall, in the latter. In both inflationary theories the entire visible Universe is produced from a single bubble. In both the crucial period of inflation is that which occurs \emph{after} bubble nucleation. In new inflation the delay between bubble nucleation and reheating is given by the slow rolling of $\phi(x)$ down the flat potential, in boom and bust inflation by the steady progression of $y(x)$ towards $d/2$. Pending a precise classical solution for the growth of a finite $\bar{R}$ bubble, we conjecture, by analogy with the equation for $\phi(x)$ in new inflation, that the unperturbed trajectory of the domain wall may be approximated as
\begin{equation}
\frac{1}{\sinh^3 H t} \frac{\partial}{\partial t} \Bigl( \sinh^3 H t \frac{\dot{{y}} }{ \sqrt{1 - \dot{{y}}^2} } \Bigl) \sim \frac{4}{\bar{R}} \ \ \ .
\end{equation} 
(For $Ht \ll 1$ this reduces to the known result for bubble growth in Minkowski space.) The wall rapidly achieves its Hubble-friction-limited subluminal terminal velocity
\begin{equation}
\dot{{y}} \sim \left[ 1 + \left({3 H \bar{R}}/{4} \right)^2 \right]^{-\frac{1}{2}} . \label{speed}
\end{equation} 
In both new inflation and boom and bust inflation the six SO(3,1) symmetries of the nucleated bubble (rotations and ``boosts'') become the six symmetries of the reheated FRW Universe (rotations and ``translations''); in both cases all six symmetries are to be weakly broken to produce CMB anisotropies, large scale structure, and, eventually, us. In both cases this breaking may be done by quantum fluctuations [fluctuations in $\phi(x)$ for new inflation, fluctuations in $y(x)$ for boom and bust inflation] that are stretched to classicality by the exponential expansion of the inflationary phase. 
Pending a precise quantum treatment of the growing bubble wall, we conjecture, by analogy with $\delta \phi_{\bf k} \sim H$ and the results of \cite{gv,gt}, that
\begin{equation}
\delta y_{\bf k} \sim \gamma^{-1} \sigma^{-1/2} H \ ,
\end{equation}
where $\sigma$ is the bubble wall tension and $\gamma^{-1} \equiv \sqrt{1 - \dot{y}^2}$ accounts for the length contraction of the perturbations. The symmetric fluctuations in the positions of the walls distort the smooth hyperboloid of Eq.~(\ref{bighyperboloid}) by a curvature perturbation 
\begin{equation} \label{per}
\mathcal{R}_{\bf k} = - \left[ \frac{H \delta y_{\bf k}}{\dot{y}} \right] _{t = t_{*}} \ ,
\end{equation} 
where $t_{*}$ is the time at which the mode with wavenumber $\bf k$ crossed the horizon. 
Since Eq.~(\ref{per}) is $\bf k$ independent, the perturbations are scale invariant. (This differs from the strongly-tilted spectrum found for bubbles colliding in Minkowski spacetime \cite{gt} as here Hubble friction keeps $\gamma$ fixed.) Since the (relativistic) kinetic term for $y$ is non-quadratic, the perturbations are non-Gaussian (as in DBI inflation \cite{dbi}).

We will return to these questions in a future publication. 

\acknowledgements 
Thanks to Raphael Bousso, Cliff Burgess, Kurt Hinterbichler, Lam Hui, Dan Kabat, John March-Russell, Alberto Nicolis, Sarah Shandera, Stephen Shenker, and Erick Weinberg for useful feedback. My work is supported by an NSF Graduate Research Fellowship.

 \end{document}